# Stable and Unstable Accretion Flows
## with Angular Momentum near a Point Mass

Dongsu Ryu[1,2], Garry L. Brown[3], Jeremiah P. Ostriker[1], and Abraham Loeb[4,5]

[1]Princeton University Observatory, Peyton Hall
Princeton, NJ 08544

[2]Department of Astronomy and Space Science, Chungnam National University
Daejeon, 305-764, (South) Korea

[3]Department of Mechanical and Aerospace Engineering, Princeton University
Princeton, NJ 08544

[4]Institute for Advanced Study
Princeton, NJ 08540

[5]Center for Astrophysics, 60 Garden Street
Cambridge, MA 02138

# ABSTRACT

The properties of axisymmetric accretion flows of cold adiabatic gas with zero total energy in the vicinity of a Newtonian point mass are characterized by a single dimensionless parameter, the thickness of incoming flow. In the limit of thin accretion flows with vanishing thickness, we show that the governing equations become self-similar, involving no free parameters. We study numerically thin accretion flows with finite thickness as well as those with vanishing thickness. Mass elements of the incoming flow enter the computational regime as thin rings. In the case with finite thickness, after a transient period of initial adjustment, an almost steady-state accretion shock with a small oscillation amplitude forms, confirming the previous work by Molteni, Lanzafame, & Chakrabarti (1994). The gas in the region of vorticity between the funnel wall and the accretion shock follows closed streamlines, forming a torus. This torus, in turn, behaves as an effective barrier to the incoming flow and supports the accretion shock which reflects the incoming gas away from the equatorial plane. The postshock flow, which is further accelerated by the pressure gradient behind the shock, goes through a second shock which then reflects the flow away from the symmetry axis to form a conical outgoing wind. As the thickness of the inflowing layer decreases (or if the ratio of the half thickness to the distance to the funnel wall along the equatorial plan is smaller than $\sim 0.1$), the flow becomes unstable. In the case with vanishing thickness, the accretion shock formed to stop the incoming flow behind the funnel wall oscillates quasi-periodically with an amplitude comparable to the thickness. The structure between the funnel wall and the accretion shock is destroyed as the shock moves inwards toward the central mass and re-generated as it moves outwards. We suggest a possible explanation for the instability. The phenomenon may be related to the quasi-periodic oscillations observed in accreting galactic sources.

*Subject headings:* Accretion - Black Holes - Hydrodynamics - Numerical Methods - Shock Waves

## 1. INTRODUCTION

The gravitational interaction between a compact object and the surrounding gas has been the topic of many studies. Bondi (1952) discussed the spherical adiabatic accretion of gas towards a stationary point mass, and found that at a steady-state the accretion rate is given by

$$\dot{M}_{\rm acc} = 4\pi\alpha \frac{(GM)^2 \rho_\infty}{c_s^3}, \qquad (1.1)$$

where $G$ is the gravitational constant, $M$ is the mass of the object, and $\rho_\infty$ and $c_s$ are the density and sound speed of the unperturbed gas. Here, $\alpha$ is a numerical constant of order unity, which is not uniquely determined by the steady-state solution. Due to its implicit assumptions of non-relativistic treatment and spherical symmetry, Bondi's solution is valid only for flows that have a free-fall velocity smaller than the speed of light, $v_{ff} \ll c$, and a rotational velocity smaller than the free-fall velocity at large radii by a factor of $\sim v_{ff}/c \ll 1$.

In the context of more realistic astrophysical applications, Bondi's solution can be extended to include the motion of the object relative to a surrounding gas or the non-spherical accretion of gas with a finite amount of angular momentum. The accretion onto a moving object was first studied by Hoyle and collaborators (see Bondi & Hoyle 1944 and references therein). They found that the accretion rate at the steady-state of cold gas with a negligible pressure onto an object moving with the velocity $v_\infty$ relative to the gas is given by

$$\dot{M}_{\rm acc} = 4\pi\alpha' \frac{(GM)^2 \rho_\infty}{v_\infty^3}, \qquad (1.2)$$

where $\alpha'$ is again a numerical constant of order unity. However, more recent numerical simulations showed that a supersonic accretion flow with a high Mach number has a strong tendency to form a dome-like shock in front of the accreting object which is unstable (see Anzer, Börner, & Monaghan 1987; Shima *et al.* 1985; Matsuda *et al.* 1991; Koide, Matsuda, & Shima 1991; Ishii *et al.* 1993; Ruffert 1994; Ruffert & Arnett 1994 and references therein). In some simulations the shock is generated and destroyed quasi-periodically, while in other simulations the shock oscillates quasi-periodically. In all cases, the flow becomes non-axisymmetric. A steady structure forms either if the accreting object is so large that the shock does not detach from its surface, or if the accretion flow has a low Mach number, or if the numerical resolution is too coarse to correctly treat the quasi-periodic structural variations.

Studies of non-spherical accretion of gas with angular momentum demonstrated the variety and complexity of structures forming around the central object (see Hawley, Smarr, & Wilson 1984a, 1984b; Hawley 1986; Hawley, & Smarr 1986; Eggum, Coroniti, & Katz 1985, 1987, 1988; Chakrabarti 1989; Chakrabarti, & Molteni 1993; Molteni, Lanzafame, & Chakrabarti 1994 and references therein). Hawley and collaborators argued that the adiabatic accretion of cold gas onto a black hole can be characterized by two parameters: the scale height and the specific angular momentum of the incoming flow. As the flow approaches the central object, it encounters an increasing centrifugal force and the gas is deflected away before reaching the black hole horizon if its angular momentum is large. In numerical simulations, Hawley and collaborators found that a travelling accretion shock forms and moves outward in the disk, but Chakrabarti and collaborators argued that a stable standing shock forms if the simulations run longer. However, Eggum and collaborators showed that no shock can be formed in a viscous accretion (Eggum, Coroniti, & Katz 1985, 1987, 1988). In all cases, the accretion would take place in a thick or a thin disk configuration around the black hole and would be accompanied by a conical outflowing wind depending on the values of the scale height and the specific angular momentum of the incoming flow.

As a simple extension of Bondi's solution, we consider in this paper the properties of thin accretion flows with angular momentum near a Newtonian point mass. It could be classified as the thin flow with angular momentum larger than the marginally bounded value according to the classification by Hawley and his collaborators (see Hawley & Smarr 1986). The purpose of this work is to understand the basic physical characteristics of these rotating accretion flows as well as to provide ground for future studies that will add more realistic physical details such as accretion to the central object, viscosity, energy transport and general relativistic effects. The paper is organized as follows. In §2, we describe the problem by defining the normalization units and writing the equations in these units. In §3, we expand the equations in powers of a small parameter, *i.e.*, the inflow thickness in units of the distance to the funnel wall, and show that the scaled equations become self-similar in the limit of vanishing thickness. In §4, we analyze the stead-state flow structure around the central object obtained from numerical simulations in the case with finite inflow thickness. The calculations discussed in this section show a good qualitative agreement with similar calculations done recently by Molteni, Lanzafame, & Chakrabarti (1994). In §5, we describe the numerical simulations of the thin accretion flows with vanishing inflow thickness, which show the unstable behavior. In §6, we suggest a possible explanation for the unstable behavior of the thin accretion flow with vanishing thickness based on the *one and a half* dimensional model. Finally, we discuss briefly the astrophysical implications of this work in §7. Appendix A describes the numerical scheme used to solve the hydrodynamic equations.

## 2. FORMULATION OF THE PROBLEM

As the simplest case, we consider the adiabatic hydrodynamics of axisymmetric flows of gas under the Newtonian gravitational field of a point mass $M$ located at the center in cylindrical coordinates $[r, \theta, z]$. We assume that at infinity the gas pressure is negligible, the net energy per unit mass vanishes, and the specific angular momentum is $j$. The gas mass in the accreting flow is assumed to be much smaller than the mass of the central object, so the self-gravity of the gas is negligible compared to the gravity of the central object. The flow is then characterized by two parameters, $M$ and $j$, that can be combined with the gravitational constant to give units of length and time,

$$[L] = \frac{j^2}{GM}, \tag{2.1}$$

$$[T] = \frac{j^3}{(GM)^2} \tag{2.2}$$

The hydrodynamic conservation equations can then be normalized by these units. The gas density scales out, so that any density can be used as its normalization unit (*i.e.*, no unit of mass is required). The resulting normalized equations in cylindrical coordinates are,

$$\frac{\partial \rho}{\partial t} + \frac{1}{r}\frac{\partial (r\rho v_r)}{\partial r} + \frac{\partial (\rho v_z)}{\partial z} = 0, \tag{2.3}$$

$$\frac{\partial v_r}{\partial t} + v_r\frac{\partial v_r}{\partial r} + v_z\frac{\partial v_r}{\partial z} + \frac{1}{\rho}\frac{\partial p}{\partial r} = \frac{v_\theta^2}{r} - \frac{r}{\left(r^2+z^2\right)^{\frac{3}{2}}} \tag{2.4}$$

$$\frac{\partial v_\theta}{\partial t} + v_r\frac{\partial v_\theta}{\partial r} + v_z\frac{\partial v_\theta}{\partial z} = -\frac{v_r v_\theta}{r}, \tag{2.5}$$

$$\frac{\partial v_z}{\partial t} + v_r\frac{\partial v_z}{\partial r} + v_z\frac{\partial v_z}{\partial z} + \frac{1}{\rho}\frac{\partial p}{\partial z} = -\frac{z}{\left(r^2+z^2\right)^{\frac{3}{2}}} \tag{2.6}$$

$$\frac{\partial p}{\partial t} + v_r\frac{\partial p}{\partial r} + v_z\frac{\partial p}{\partial z} + \gamma\frac{p}{r}\frac{\partial (rv_r)}{\partial r} + \gamma p\frac{\partial v_z}{\partial z} = 0, \tag{2.7}$$

where $p$, $\rho$ and $\boldsymbol{v}$ are the normalized pressure, mass density and velocity of the gas. The thermal and total energies are

$$u = \frac{p}{(\gamma-1)}, \tag{2.8}$$

and

$$E = \frac{p}{(\gamma-1)} + \frac{\rho(v_r^2 + v_\theta^2 + v_z^2)}{2}. \tag{2.9}$$

Apart from shocks, the equations imply that entropy is conserved, *i.e.*,

$$\frac{D(p/\rho^\gamma)}{Dt} = 0. \tag{2.10}$$

Note that, in the case of axisymmetric flow without viscosity, the equation for the azimuthal momentum, Eq. (2.5), states a conservation of angular momentum

$$\frac{d(rv_\theta)}{dt} = 0, \tag{2.11}$$

where $rv_\theta = 1$ for accreting gas in our normalized unit.

The assumption of zero net energy per unit mass *at infinity* in the accreting gas implies the constraint for steady flows that the enthalpy plus the kinetic and potential energies per unit mass must be zero, *i.e.*,

$$\frac{v_r^2}{2} + \frac{v_z^2}{2} + \frac{c_s^2}{\gamma-1} = \frac{1}{\sqrt{r^2+z^2}} - \frac{1}{2r^2}, \tag{2.12}$$

where $c_s = \sqrt{\gamma p/\rho}$ is the sound speed and the relation $v_\theta = 1/r$ has been used.

As gas approaches the central object, it cannot get closer to the symmetry axis than the radius of the *funnel wall*, a barrier formed due to the centrifugal force. The role of the funnel wall in the dynamics of accretion flows was discussed by Hawley & Smarr (1986). In the case of Newtonian gravity, the location of the funnel wall for the gas with zero net energy can be found from the energy constraint (2.12). The position of the funnel wall is defined as the surface in the $r$-$z$ plane along which the right hand side of (2.12) vanishes, *i.e.*,

$$2r^2 = \sqrt{r^2 + z^2}. \tag{2.13}$$

Inside the funnel wall the right hand size of (2.12) becomes negative, which implies that the kinetic energy of $r$ and $z$ motion plus the enthalpy of the gas become negative. Hence, the centrifugally supported gas with zero net energy cannot move inside of the funnel wall.

Even though the above normalized equations do not contain any preferred time or length scale, the funnel wall introduces a length scale, *i.e.*, the distance ($r_f$) to the funnel wall from the central mass point along the equatorial plane ($z = 0$). In our dimensionless unit, $r_f = 1/2$. Hence, the flow structure around a Newtonian point mass is determined by a single parameter describing the geometry of the incoming flow, *i.e.*, the ratio of the half thickness of incoming flow ($h$) to the distance to the funnel wall, $\epsilon \equiv h/r_f$. For a relativistic black hole with a finite horizon size, there are two length scales in the problem, the distance to the funnel wall and the Schwarzschild radius. Hence, the flow structure is determined by two parameters, *i.e.*, the geometry and the specific angular momentum of incoming flow, as pointed by Hawley and collaborators.

## 3. EQUATIONS IN THE LIMIT OF VANISHING INFLOW THICKNESS

In the limit of vanishing thickness of accreting inflow ($\epsilon \to 0$), the gas along the equatorial plane approaches very close to the funnel wall. Thus, we focus on the local dynamics of gas in the vicinity of funnel wall. In the following, we derive the equations suitable to study this local dynamics.

At the funnel wall along the equatorial plane ($r = r_f$ and $z = 0$), the azimuthal velocity of the accreting gas approaches $v_{\theta f} = 2$ and the angular velocity approaches $\Omega_f = 4$ in our dimensionless units. First, we define cylindrical coordinates rotating with the accreting gas at the funnel wall as follows

$$r' = r, \qquad \theta' = \theta - \Omega_f t, \qquad z' = z,$$

$$v'_r = v_r, \qquad v'_\theta = v_\theta - \Omega_f r, \qquad v'_z = v_z. \tag{3.1}$$

Then, the equations (2.3) to (2.7) become

$$\frac{\partial \rho'}{\partial t} + \frac{1}{r'}\frac{\partial \left(r'\rho' v'_r\right)}{\partial r'} + \frac{\partial \left(\rho' v'_z\right)}{\partial z'} = 0, \tag{3.2}$$

$$\frac{\partial v'_r}{\partial t} + v'_r \frac{\partial v'_r}{\partial r'} + v'_z \frac{\partial v'_r}{\partial z'} + \frac{1}{\rho'} \frac{\partial p'}{\partial r'} = \frac{v'^2_\theta}{r'} - \frac{r'}{\left(r'^2 + z'^2\right)^{\frac{3}{2}}} + 2\Omega_f v'_\theta + \Omega_f^2 r', \quad (3.3)$$

$$\frac{\partial v'_\theta}{\partial t} + v'_r \frac{\partial v'_\theta}{\partial r'} + v'_z \frac{\partial v'_\theta}{\partial z'} = -\frac{v'_r v'_\theta}{r'} - 2\Omega_f v'_r, \quad (3.4)$$

$$\frac{\partial v'_z}{\partial t} + v'_r \frac{\partial v'_z}{\partial r'} + v'_z \frac{\partial v'_z}{\partial z'} + \frac{1}{\rho'} \frac{\partial p'}{\partial z'} = -\frac{z'}{\left(r'^2 + z'^2\right)^{\frac{3}{2}}}, \quad (3.5)$$

$$\frac{\partial p'}{\partial t} + v'_r \frac{\partial p'}{\partial r'} + v'_z \frac{\partial p'}{\partial z'} + \gamma \frac{p'}{r'} \frac{\partial \left(r' v'_r\right)}{\partial r'} + \gamma p' \frac{\partial v'_z}{\partial z'} = 0. \quad (3.6)$$

We further define a pseudo-Cartesian coordinate system which extends over a small region around the funnel wall with scales comparable to the thickness of the incoming flow:

$$x = r' - r'_f, \qquad y = r'_f \left(\theta' - \theta_o\right), \qquad z = z',$$

$$v_x = v'_r, \qquad v_y = \frac{r'_f}{r'} v'_\theta, \qquad v_z = v'_z. \quad (3.7)$$

We consider a situation where

$$\frac{x}{r_f} \sim \frac{y}{r_f} \sim \frac{z}{r_f} \sim \epsilon \ll 1, \quad (3.8)$$

and

$$\frac{v_x}{r_f \Omega_f} \sim \frac{v_z}{r_f \Omega_f} \sim \sqrt{\epsilon}$$

$$\frac{v_y}{r_f \Omega_f} \sim \epsilon. \quad (3.9)$$

Expanding the equations (3.2) to (3.6) with respect to $\epsilon$ and keeping only the lowest non-zero power of $\epsilon$, then, the resulting equations are

$$\frac{\partial \rho}{\partial t} + \frac{\partial (\rho v_x)}{\partial x} + \frac{\partial (\rho v_z)}{\partial z} = 0, \quad (3.10)$$

$$\frac{\partial v_x}{\partial t} + v_x \frac{\partial v_x}{\partial x} + v_z \frac{\partial v_x}{\partial z} + \frac{1}{\rho} \frac{\partial p}{\partial x} = 4, \quad (3.11)$$

$$\frac{\partial v_y}{\partial t} + v_x \frac{\partial v_y}{\partial x} + v_z \frac{\partial v_y}{\partial z} = -2\Omega_f v_x, \quad (3.12)$$

$$\frac{\partial v_z}{\partial t} + v_x \frac{\partial v_z}{\partial x} + v_z \frac{\partial v_z}{\partial z} + \frac{1}{\rho} \frac{\partial p}{\partial z} = 0, \quad (3.13)$$

$$\frac{\partial p}{\partial t} + v_x \frac{\partial p}{\partial x} + v_z \frac{\partial p}{\partial z} + \gamma p \frac{\partial v_x}{\partial x} + \gamma p \frac{\partial v_z}{\partial z} = 0. \quad (3.14)$$

Note that the equation (3.12) for $v_y$, which provides for the conservation of angular momentum, is decoupled from the rest of the equations as pointed in the previous section. Hence, the dynamical time scale for the above equations is

$$\tau_d \sim \frac{x}{v_x} \sim \frac{z}{v_z} \sim \frac{\sqrt{\epsilon}}{\Omega_f}, \qquad (3.15)$$

which is much smaller than the orbital time scale $1/\Omega_f$. Also note that the equation for the vertical velocity does not contain any contribution due to gravity from the central object.

The above equations are similar to but different from the widely used *shearing sheet* equations (Goldreich, & Lynden Bell 1965; Julian, & Toomre 1965; Shu 1974) due to the different scaling assumed for $v_x$ and $v_z$ with $\epsilon$ in Eq. (3.9). The resulting equations can be considered as those which describe a two-dimensional flow in $x - z$ plane moving under a constant force in the direction of positive $x$.

In this local coordinate system, the equation for the energy constraint in Eq. (2.12) becomes

$$\frac{v_x^2}{2} + \frac{v_z^2}{2} + \frac{c_s^2}{\gamma - 1} = 4x. \qquad (3.16)$$

Since we expanded the equations around the funnel wall, the origin moved from the position of the central mass point to the position of the funnel wall along the equatorial plane. Hence, the length scale introduced to the problem in §2, the distance to the funnel wall, drops out of the consideration. So the flow in this limit does not contain any free parameter, and can be scaled in a *self-similar* way to different scales of length, time, and gas mass.

## 4. STABLE SOLUTIONS FOR THE FLOWS WITH FINITE INFLOW THICKNESS

Before turning to the self-similar case, we examine the solution to the normalized equations in §2. The equations were solved numerically in a computational box which occupies one quadrant of the $r$-$z$ plane with $0 \leq r \leq 4$ and $0 \leq z \leq 4$, using a two-dimensional hydrodynamic code based on the TVD scheme which is described in Appendix A. The code was designed to be able to handle the general case where the angular momentum is not necessarily conserved for the future expansions of the work, even though the angular momentum is conserved in the problem considered in this paper. The point mass is located at the origin. Note that in our normalized units, $r = 1/2$ is the location of the funnel wall on the equatorial plane and $r = 1$ is the point where the centrifugal force balances the gravitational force.

The incoming flow, which enters the computational box along the equatorial plane through the boundary located at $r^* = 4$, has $\gamma = 5/3$, $\rho = 1$, $v_\theta = 1/4$, $v_z = 0$, and $c_s/v_r = 1/10$. The radial velocity and sound speed of the incoming flow are determined by equation (2.12). We simulated four cases with different half-thicknesses of incoming flow: $h_{\text{in}} = 1/8$ (Run 1), 2/8 (Run 2), 3/8 (Run 3), and 4/8 (Run 4). Note that our dimensionless parameter is $\epsilon \equiv h_{\rm in}/r_f = 2h_{\rm in}$. All runs used an equally spaced grid with $128 \times 128$ cells.

For the boundary condition of the outgoing flow, it is appropriate to use a non-reflecting, radiative condition. Instead, we used a simple continuous condition in which all the derivatives of fluid quantities were set to be zero at the boundaries. We found that even with the simple continuous condition, reflection at the boundary was kept at a minimum and did not affect the structure formed along the equatorial plane. Along the symmetry axis ($r = 0$) and the equatorial plane ($z = 0$), the usual reflecting condition was used.

The gas which enters the computational box as a thin inflow is decelerated in the region between the funnel wall and the balance point for radial forces and is eventually reflected. The reflected gas, then, leads to the formation of an accretion shock. The shock heats the gas and the hot gas in turn pushes the shock out to a large radius until the pressure of the hot gas finds a balance with the ram pressure of the steady stream of the cold incoming gas. After a few oscillations around the steady configuration, a steady torus-like structure forms between the funnel wall and the shock.

In Fig. 1, we plotted the time evolution of positive (kinetic plus internal) energy (total energy inside the computational box divided by total volume), $E$, and bremsstrahlung emission (total emission divided by total volume), $\dot{E}_{\rm brem}$, in the four simulations. Since the gas density is arbitrary, we characterize the bremsstrahlung emissivity simply by

$$j_{\rm brem} = \rho^{3/2} p^{1/2}. \qquad (4.1)$$

The free-fall stage ends around $t \sim 5$ forming the accretion shock and an approximate steady-state is obtained after $t \sim 50$.

Once the system relaxes to an approximate steady-state, the overall structure remains unchanged even though there are still changes in the detailed structure. As an example of a steady-state, we present the velocity field at the end of Run 4 in Fig. 2. Fig. 3 shows the gray scale images of the density, pressure, entropy, and Mach number at the end of the same simulation. The Mach number was computed using the full velocity vector,

$$\mathcal{M} = \frac{\sqrt{v_r^2 + v_\theta^2 + v_z^2}}{c_s}. \qquad (4.2)$$

At the steady-state, the toroidal volume between the funnel wall and the standing accretion shock confines two or more vortices. The gas trapped in the vortices follows roughly closed loops in the $r - z$ plane, as indicated by the dotted lines in Fig. 2. The two large vortices in the toroidal volume just behind the shock behave as if they were an actual physical barrier to the incoming flow and reflect the incoming stream of gas away from the equatorial plane through the oblique shock. The smaller vortices in between the large vortices and the funnel wall can be interpreted as secondary vortices that form due to the effects of the large vortices. With better resolution, we would expect to observe even smaller vortices. The entropy plot in Fig. 3c indicates that the gas trapped in the vortices and the gas reflected at the shock have different entropies, implying that they went through different physical histories. The trapped gas in the vortices has a higher entropy, indicating it was shocked at the earlier adjustment period during which the shock was closer to the funnel wall and stronger. The separation between the phases is preserved at all times.

Note that in Fig. 2 the innermost fluid elements reach to $r < r_f = 1/2$. This is not caused by numerical errors but is due to the fact that pressure force can push the gas into a region that would be forbidden to freely falling gas.

It is interesting to note that for an oblique shock supported by a barrier, the shock angle $\theta_{\text{shock}}$ and the flow deflection angle $\theta_{\text{def}}$ are related. In the limit of very high Mach number, the two angles satisfy

$$\tan\theta_{\text{def}} = \frac{\sin 2\theta_{\text{shock}}}{\gamma + \cos 2\theta_{\text{shock}}}, \tag{4.3}$$

(see, *e.g.*, Liepmann & Roshko 1957). From Fig. 2 we estimate $\theta_{\text{shock}} = 42°$ and $\theta_{\text{def}} = 29°$, which satisfy the above relation for $\gamma = 5/3$. (Here the two angles were estimated by including the azimuthal velocity, $v_\theta$.)

The gas reflected by the shock is further accelerated by the pressure gradient behind the shock, as it moves away from the equatorial plane. Then, it undergoes another shock when the fluid elements, which have entered through the outer boundary very close to the equatorial plane, reach the funnel wall above the torus-like structure. By the continuation of the shock, even though the fluid elements which enter far from the equatorial plane do not reach the funnel wall, they still undergo the shocking process. This second shock reflects the flow away from the symmetry axis, forming a conical outgoing wind.

Two additional vortices persist in Fig. 2 just above and below the incoming flow. They form in the regions trapped by the trajectories of the fluid elements which enter along the top and bottom edges of the incoming flow. The gas contained in those vortices is mostly initial background gas and also follows roughly closed loops.

The picture of the flow around the accretion shock differs from the work by Hawley and collaborators (see Hawley 1986; Hawley & Smarr 1986) but agrees with that by Chakrabarti and collaborators (see Molteni, Lanzafame, & Chakrabarti 1994), in the fact that the vortices of trapped gas, which allow the flows to be deflected by the accretion shock, have been identified. In addition, we have found a second shock formed above and below the torus-like structure, which reflects the flow away from the symmetry axis. These differences could be accounted for by the differences in numerical resolution achieved and in the numerical schemes used in the various works. Since we have used a code which is optimized to capture shocks and have focused on the structure just around the central object, we have been able to get the second shock as well as the standing accretion shock. Independent studies by Eggum, Coroniti, & Katz (1985, 1988) also found a vorticity structure similar to ours. But they did not resolve shocks because of the viscosity prescription included in their simulations.

In Fig. 4, the velocity fields at the end of four simulations are plotted along with the line following the funnel wall. The qualitative features remain similar in these simulations, although they are characterized by different thickness. However, as the dimensionless parameter increases in the sequence (a) → (d) from $\epsilon = 1/4$ to $\epsilon = 1$, the outflow cone becomes wider, the accretion shock moves back from the funnel wall, the toroidal volume between the funnel wall and the shock increases, and the number of trapped vorticities increases.

In Fig. 5 we show four quantities, namely, (a) the bremsstrahlung emission (total emission from the computational box divided by total volume), $\dot{E}_{\rm brem}$, (b) the angle of the conical outflowing wind

$$\theta_{\rm out} \equiv \int_0^\pi \theta \frac{d\dot{M}_{\rm out}}{d\theta} \sin\theta d\theta \bigg/ \int_0^\pi \frac{d\dot{M}_{\rm out}}{d\theta} \sin\theta d\theta, \qquad (4.4)$$

(c) the position of the accretion shock along the equatorial plane, $r_s$, in the unit of the rotational support radius where the centrifugal force balances the gravitational force, and (d) the mass trapped within the shock annulus

$$m_A = \int_{-\infty}^{\infty} dz \int_0^{r_s} \rho 2\pi r dr \qquad (4.5)$$

with $\rho = 1$, as a function of $\epsilon$. Here, $d\dot{M}_{\rm out}/d\theta$ is the mass escape rate of the outflow wind per unit angle at angle $\theta$ through a sphere of radius equal to $R = r^* = 4$. From the plots, we see that the distance between the funnel wall and the accretion shock is linearly proportional to $\epsilon$ with

$$\frac{r_s}{r_f} - 1 \approx 4.6\epsilon, \qquad (4.6)$$

which is consistent with the self-similar behavior described in §3. The angle of the conical outflowing wind is also approximately linearly proportional to $\epsilon$. Note that although the inflow mass which enters the computational box during a unit time period is linearly proportional to $\epsilon$ ($\sim 16.4\epsilon$ in units), the mass trapped within the shock annulus is not.

## 5. UNSTABLE SOLUTIONS FOR THE FLOWS WITH VANISHING INFLOW THICKNESS

The equations derived in §3 for the case $\epsilon \to 0$ were solved numerically with the two-dimensional plane-parallel TVD code. The one-dimensional plan-parallel TVD code was described in the original paper about the TVD scheme by Harten (1983). We used the multi-dimensional version which was expanded from the one-dimensional code using the Strang-type operator splitting (see Ryu et al.(1993) for more details).

The computational box encompasses a region in the $x - z$ plane with $-0.4 \leq x \leq 0.6$ and $0 \leq z \leq 1$. Note that the origin corresponds to the position of the funnel wall along the equatorial plane. The flow enters the computational box along the equatorial plane through the boundary located at $x^* = 0.6$ with half thickness $1/32$ and has $\gamma = 5/3$, $\rho = 1$, $v_z = 0$, and $c_s/v_x = 1/10$. Then the $x$-velocity and sound speed are determined by the equation (3.16). For the boundary condition of the outgoing flow, we used the simple continuous condition. Along the equatorial plane with $z = 0$, we used the usual reflecting condition.

In Fig. 6, the time evolution of positive (kinetic plus internal) energy per unit volume and bremsstrahlung emission per unit volume were plotted. The plot shows two different phases, the quiet phase (*e.g.*, around $t \sim 70$) with small amplitude fluctuations in the changes of the bremsstrahlung emission and the active phase (*e.g.*, around $t \sim 90$) with large amplitude fluctuations.

In the quiet phase, the position of the accretion shock does not vary very much nor does the flow structure. Fig. 7 shows the flow velocity field at the several different epochs during the quiet phase. In this plot, we do not see any obvious vorticity behind the shock and also we do not see the second shock described in the previous section. The gas which enters the computational box is decelerated and eventually stops, forming an accretion shock before the origin (the funnel wall). It is then simply deflected and becomes the outgoing wind without accompanying a second shock.

However, in the active phase, the position of the shock is not fixed but it rather oscillates quasi-periodically between $0.1 \lesssim x \lesssim 0.4$ with a time scale of $\sim 2.5$. In Fig. 8, the velocity field at several different epochs during the active phase is plotted. As the shock moves inwards or outwards, the structure between the origin and the shocks is destroyed and re-generated. Depending on the position of the shock, the conical structure of the outgoing flow also changes significantly. In the next section, we describe a possible origin of this unstable behavior.

## 6. POSSIBLE EXPLANATION FOR THE UNSTABLE ACCRETING FLOW

Let us consider an *one*-dimensional cylindrical flow, which enters the domain of consideration through a boundary at $r^*$. For a steady flow, the conservations of mass and energy give

$$\begin{aligned} r\rho v_r &= \text{constant} \\ &= r^* \rho^* v_r^*, \end{aligned} \tag{6.1}$$

and

$$\frac{c_s^2}{\gamma - 1} + \frac{1}{2}v_r^2 + \frac{1}{2r^2} - \frac{1}{r} = \text{constant} \tag{6.2}$$
$$= 0.$$

In the second relation, $1/2r^2$ is the kinetic energy due to the azimuthal velocity and the vertical velocity was assumed to zero. The flow was assumed to have zero net energy per unit mass at infinity. If the flow is isentropic without shocks, the entropy conservation gives

$$\frac{p}{\rho^\gamma} = \text{constant}$$
$$= \frac{p^*}{\rho^{*\gamma}}. \tag{6.3}$$

Here, the flow quantities were normalized with the same normalization units as in §2. $\rho^*$, $p^*$, and $v_r^*$ are the density, pressure, and radial velocity at $r^*$. In the followings, the quantities with superscript $^*$ are those defined at the boundary $r^*$. By combining the above three relations, the distributions of the radial Mach number $M_r \equiv v_r/c_s$ and pressure as a function of the radius $r$ are calculated, once the flow quantities, $v_r^*$, $c_s^*$, and $p^*$, at the boundary $r^*$ are given:

$$\frac{c_s^{*2}}{\gamma - 1} + \frac{v_r^{*2}}{2}\left(\frac{M_r}{M_r^*}\right)^2 = \left(\frac{r}{r^*}\right)^{2\frac{\gamma-1}{\gamma+1}} \left(\frac{M_r}{M_r^*}\right)^{2\frac{\gamma-1}{\gamma+1}} \left(\frac{1}{r} - \frac{1}{2r^2}\right), \tag{6.4}$$

and

$$\frac{p}{p^*} = \left(\frac{r^*}{r}\right)^{\frac{2\gamma}{\gamma+1}} \left(\frac{M_r^*}{M_r}\right)^{\frac{2\gamma}{\gamma+1}}. \tag{6.5}$$

To be complete, the above solution for a steady flow requires a sink around the origin to absorb the inflow.

Fig. 9 shows the distributions of the radial Mach number and pressure of an one-dimensional steady isentropic flow, which enters through a boundary located at $r^* = 4$ with $\rho^* = 1$, $v_\theta^* = 1/4$, $v_z^* = 0$, $M_r^* = 10$, and $\gamma = 5/3$. The above inflow parameters are the same as those of Run 1 - Run 4 described in §4. As the supersonic flow approaches to the origin, the radial Mach number increases and reaches a maximum at $r = 2/(3 - \gamma) = 1.5$. Then it decreases until $M_r = 1$, where the above solution breaks down since a supersonic flow can not change to a subsonic flow without involving a shock. The position of the maximum radial Mach depends only on the value of $\gamma$. On the other hand, the pressure continues to increase until the flow ceases to be valid at $M_r = 1$. This solution describes reasonably well the flow considered in §4 in the region between the boundary and the accretion shock, where the vertical motion is unimportant, as also pointed by Chakrabarti (1989).

If the incoming flow is subsonic, the radial Mach number decreases as the flow approaches to the origin reaching a minimum at $r = 2/(3 - \gamma)$ and then starts to increase. The pressure increases first and then decreases. The maximum point of the pressure generally does not coincide with the minimum point of the radial Mach number, and depends on the parameters of the incoming flow.

If a shock exists at $r_s$ in an one-dimensional steady flow, the flow upstream of the shock ($r_s < r \leq r^*$) is described by a solution with $v_{r1}^*$, $c_{s1}^*$, and $p_1^*$ at the boundary $r^*$ and the flow downstream of the shock ($r < r_s$) is described by another solution with $v_{r2}^*$, $c_{s2}^*$, and $p_2^*$ at the boundary $r^*$. Once the parameters of the upstream flow are given and the shock position is fixed, the shock Mach number, $M_s$, is calculated from the equation (6.4) and the parameters for the downstream flow are calculated as follows. The entropy change across the shock is a function of the shock Mach number

$$\left(\frac{p_2}{p_1}\right)\left(\frac{\rho_1}{\rho_2}\right)^\gamma = f(M_s), \tag{6.6}$$

and the entropy conservation dictates

$$\left(\frac{p_2}{p_1}\right)\left(\frac{\rho_1}{\rho_2}\right)^\gamma = \left(\frac{p_2^*}{p_1^*}\right)\left(\frac{\rho_1^*}{\rho_2^*}\right)^\gamma. \tag{6.7}$$

Here, the quantities with subscript 1 are for the upstream flow and those with 2 are for the downstream flow. From the shock jump condition of mass flux and the mass conservation

$$\rho_1^* v_{r1}^* = \rho_2^* v_{r2}^*, \tag{6.8}$$

and from the shock jump condition of energy flux and the energy conservation

$$\frac{c_{s1}^*}{\gamma - 1} + \frac{1}{2}v_{r1}^{*2} = \frac{c_{s2}^*}{\gamma - 1} + \frac{1}{2}v_{r2}^{*2}. \tag{6.9}$$

By combining the equations (6.6) to (6.9), we get

$$f(M_s) = \left(\frac{1 + \frac{\gamma-1}{2}M_{r1}^{*2}}{1 + \frac{\gamma-1}{2}M_{r2}^{*2}}\right)^{\frac{\gamma+1}{2}} \left(\frac{M_{r2}^*}{M_{r1}^*}\right)^{\frac{\gamma+1}{2}}. \tag{6.10}$$

On the other hand, from the usual shock jump conditions, we get

$$f(M_s) = \left[1 + \frac{2\gamma}{\gamma + 1}(M_s^2 - 1)\right]\left[\frac{1 + (\gamma - 1)M_s^2}{(\gamma + 1)M_s^2}\right]^\gamma. \tag{6.11}$$

If $M_{r1}^*$ and $M_s$ is given, $M_{r2}^*$ is calculated from (6.10) and (6.11). Then, the parameters for the downstream flow are calculated from (6.8) to (6.9).

Fig. 10 shows the pressure distributions of one-dimensional steady flows with shocks located at different radii. The flow entering through a boundary at $r^* = 4$ has $\rho^* = 1$, $v_\theta^* = 1/4$, $v_z^* = 0$, $M_r^* = 10$, and $\gamma = 5/3$, same as those of Fig. 9. The upper plot compares the pressure distributions of flows with shocks at $r = 2.4$ and $r = 2.5$, in the region where the shock Mach number decreases with radius ($r > 1.5$, see Fig. 9). Here, the downstream pressure with larger shock Mach number (the flow with shock at $r = 2.4$) is larger than that with smaller shock Mach number. The lower plot compares the pressure distributions with shocks at $r = 0.9$ and $r = 1.0$, in the region where the shock Mach number increases with radius ($r < 1.5$). In this case, the downstream pressure with larger shock Mach number (the flow with shock at $r = 1.0$) is smaller. This behavior is generally true, depending only on whether the shock is located at $r > 1, 5$ or at $r < 1.5$.

For $r < 1.5$, as $r$ increases, $M_s$ increases for isentropic flows (see Fig. 9) and the entropy change across a shock increases. But the pressure at the same radius downstream of the shock falls. So if for some reason (*e.g.*, unsteady flows) the pressure downstream of the shock is above the equilibrium value at that shock location, the shock will move to a larger radius to give a larger $M_s$ relative to the incoming flow. However, the larger $M_s$ at the larger radius has a lower equilibrium pressure at the downstream radius. So the difference between given and equilibrium pressure is increased further. Thus, there is no stable configuration until the shock moves to $r > 1.5$, where $M_s$ falls with increasing $r$.

Because of the existence of the vertical motion in the postshock region, the two-dimensional numerical calculations described in the previous two sections are much more complicated than the above one-dimensional solution for steady flows. However, we think the consideration of stability is still applicable to the numerical calculations and causes the oscillatory behavior in the case with vanishing inflow thickness. But in the unstable configuration the shock does not travel up to $r > 1.5$, because the vertical motion evacuates the postshock region causing the shock to move back. Also our calculations show that in order to show a noticeable oscillatory behavior the shock Mach number should increases rather sharply with the radius. So the flows only with shock in $r \lesssim 1$ (or $\epsilon \lesssim 0.1$, see Fig. 5) become unstable.

Prior investigations of accreting sources (see references in §1) did not find the unstable solution that we have identified for one of the following reasons: the dimensionless thickness of the incoming flow $\epsilon$ is was large and the solution was in fact stable (Molteni, Lanzafame, & Chakrabarti 1994), the calculation was not continued for a sufficiently long time (Hawley 1986; Hawley & Smarr 1986), or available numerical resolution was insufficient. We are fairly confident that the phenomena found here are real and can be understood in terms of the physical mechanism described above.

## 7. ASTROPHYSICAL IMPLICATIONS

Since real astronomical accreting flows are often thought to involve very thin disks, it is tempting to identify the instability found by us with those shown in real observed systems. In fact a large fraction of the accreting astronomical x-ray sources observed carefully are known to show short time-scale variability at frequencies comparable to the orbital frequency at the rotational support radius ($r_f$ in our notation).

Of special interest are the low mass x-ray binaries which typically show quasi-periodic oscillations - the QPO phenomenon. The prevailing belief is that these involve neutron stars accreting from less mass companions, which are gradually spiraling together. The observed x-rays are powered by accretion onto the neutron star, but there is often evidence for substantial mass loss as well. A fairly recent review of the QPO phenomena with references to the earlier literature is presented by Van der Klis (1989). Beat frequencies between the neutron star rotation frequency and other natural frequencies are often invoked to explain the QPO phenomena, but it has remained as a serious problem for this scenario that the underlying fundamental rotation frequency has never been detected, even when searches were carried to quite high sensitivity.

When one performs Fourier analyses of the temporal variability in QPO sources, one finds typically in any time period a definite peak in Fourier space with a frequency corresponding to the oscillation period - hence the *quasi* appellation. The power spectrum of the normalized bremsstrahlung emission in Fig. 6

$$P = \left| \int \left( \frac{\dot{E}_{\text{brem}}}{h_{\text{in}}^2} - \left\langle \frac{\dot{E}_{\text{brem}}}{h_{\text{in}}^2} \right\rangle \right) e^{i2\pi\nu t} dt \right|^2 \tag{7.1}$$

in the time interval $25 \leq t \leq 100$ shown in Fig. 11 also indicates a similar behavior with a peak corresponding to a period of $\sim 2.5$. In calculating the power spectrum, we did not apply any windowing and smoothing.

Is the range of frequencies typically found in QPO sources - between 5 and 60 Hz - plausible for our mechanism? The instability only occurs for $\epsilon \lesssim 0.1$ and the time scale is related to the free fall time between $r_s$ and $r_f$ which is order of (true)$\sqrt{\epsilon}$ times the period at $r_f$. Or approximately,

$$\begin{aligned}\sqrt{\epsilon} \frac{j^3}{(GM)^2} &\sim \frac{1}{\nu} \\ &\sim \frac{1}{5} - \frac{1}{60} \text{ sec},\end{aligned} \tag{7.1}$$

not a totally irrelevant range of periods.

Another feature common between the instability which we have found and the QPO phenomena is the apparent switching between stable and oscillating states. In the oscillating state, we would expect accretion to increase as the shock periodically becomes close to $r_f$ and material would more easily reach the central object.

At this time, it is clearly premature to explore the analogy between the instability we have found and the QPO phenomenon or other observed variabilities much further, in light of extremely simplified physics which we have adopted. We plan in the future to allow for viscosity and accretion onto the central object as well as radiative loss processes. At that time, we can return to examine in greater detail the liability of the instability that we have found as a potential driver for the QPO phenomenon.

## ACKNOWLEDGMENTS

We thank John F. Hawley for discussions. The work by DR was supported in part by David and Lucille Packard Foundation Fellowship through Jeremy Goodman at Princeton University and in part by the Non-Directed Research Fund of the Korea Research Foundation 1993 at Chungnam National University. The work by JPO was in part supported by NSF grant AST 91-08103. The work by AL was in part supported by a W M Keck fellowship at Institute for Advanced Study.

## APPENDIX A

TVD HYDRODYNAMICS IN TWO-DIMENSIONAL CYLINDRICAL COORDINATES

We have solved numerically the dimensionless equations in §2 and §3 using an Eulerian finite difference code based on the total variation diminishing (TVD) scheme, originally developed by Harten (1983). The Harten's scheme is an explicit, second order accurate scheme which is designed to solve a hyperbolic system of the conservation equations, like the system of the hydrodynamic conservation equations. It is a nonlinear scheme obtained by first modifying the flux function and then applying a non-oscillatory first order accurate scheme to get a resulting second order accuracy. The key merit of this scheme is to achieve the high resolution of a second order accuracy while preserving the robustness of a non-oscillatory first order scheme. Also, the scheme is relatively simple to program compared to other high accuracy numerical schemes and requires less CPU time per time step for a given spatial resolution.

Harten (1983) described the application of his TVD scheme to the set of the one-dimensional hydrodynamic equations in plane-parallel geometry and presented some test results. While it is conceptually straightforward to expand the TVD code in one-dimensional plane-parallel geometry into that in two-dimensional plane-parallel or cylindrical geometry, the detailed implementation is somewhat involved in a practical sense. Since the procedure to build the TVD code in multi-dimensional plane-parallel geometry was already described in Ryu et al.(1993) along with tests, here we describe briefly the procedure to build the TVD code in two-dimensional cylindrical geometry. The purpose of this appendix is to provide a short but complete description of the procedure. For the details, e.g., why and how each step works, the choice of the values of the internal parameters, etc, refer the original reference (Harten 1983).

The equations in (2.3)-(2.7) are written in the vector form as

$$\frac{\partial \boldsymbol{q}}{\partial t} + \frac{1}{r}\frac{\partial (r\boldsymbol{F}_1)}{\partial r} + \frac{\partial \boldsymbol{F}_2}{\partial r} + \frac{\partial \boldsymbol{G}}{\partial z} = \boldsymbol{S}, \qquad (A.1)$$

where the state vector is

$$\boldsymbol{q} = \begin{pmatrix} \rho \\ \rho v_r \\ \rho v_\theta \\ \rho v_z \\ E \end{pmatrix}, \qquad (A.2)$$

the flux functions are

$$\boldsymbol{F}_1 = \begin{pmatrix} \rho v_r \\ \rho v_r^2 \\ \rho v_\theta v_r \\ \rho v_z v_r \\ (E+p)v_r \end{pmatrix} \qquad \boldsymbol{F}_2 = \begin{pmatrix} 0 \\ p \\ 0 \\ 0 \\ 0 \end{pmatrix} \qquad \boldsymbol{G} = \begin{pmatrix} \rho v_z \\ \rho v_r v_z \\ \rho v_\theta v_z \\ \rho v_z^2 + p \\ (E+p)v_z \end{pmatrix}, \qquad (A.3)$$

and the source function is

$$S = \begin{pmatrix} 0 \\ \frac{\rho v_\theta^2}{r} - \frac{\rho r}{(r^2+z^2)^{\frac{3}{2}}} \\ -\frac{\rho v_r v_\theta}{r} \\ -\frac{\rho z}{(r^2+z^2)^{\frac{3}{2}}} \\ -\frac{\rho(rv_r+zv_z)}{(r^2+z^2)^{\frac{3}{2}}} \end{pmatrix} \quad (A.4)$$

The equation of state is given by $E = p/(\gamma - 1) + \rho(v_r^2 + v_\theta^2 + v_z^2)/2$.

With the state vector, $q$, and the flux functions, $F(q) = F_1(q) + F_2(q)$ and $G(q)$, the Jacobian matrices, $A(q) = \partial F/\partial q$ and $B(q) = \partial G/\partial q$, are formed. The system of equations in (A.1) is called *hyperbolic*, if all the eigenvalues of the Jacobian matrix are real and the corresponding set of right eigenvectors is complete.

The eigenvalues and the right and left eigenvectors of $A(q)$ which were used to build the TVD code are listed below. The eigenvalues and eigenvectors of $B(q)$ are obtained by properly permuting the indices. The eigenvalues of $A(q)$ are

$$\begin{aligned} a_1 &= v_r - c, \\ a_2 &= v_r, \\ a_3 &= v_r, \\ a_4 &= v_r, \\ a_5 &= v_r + c, \end{aligned} \quad (A.5)$$

where the sound speed is $c = \sqrt{\gamma p/\rho}$. The corresponding right eigenvectors are

$$R_1 = \begin{pmatrix} 1 \\ v_r - c \\ v_\theta \\ v_z \\ H - v_r c \end{pmatrix}, \quad R_2 = \begin{pmatrix} 0 \\ 0 \\ 1 \\ 0 \\ v_\theta \end{pmatrix}, \quad R_3 = \begin{pmatrix} 1 \\ v_r \\ v_\theta \\ v_z \\ \Theta/2 \end{pmatrix}, \quad R_4 = \begin{pmatrix} 0 \\ 0 \\ 0 \\ 1 \\ v_z \end{pmatrix}, \quad R_5 = \begin{pmatrix} 1 \\ v_r + c \\ v_\theta \\ v_z \\ H + v_r c \end{pmatrix}, \quad (A.6)$$

where $H = (E + p)/\rho$ is the enthalpy and $\Theta = v_r^2 + v_\theta^2 + v_z^2$. The left eigenvectors which are orthonormal to the right eigenvector, $L_l \cdot R_m = \delta_{lm}$, are

$$L_1 = \left( \frac{(\gamma-1)\Theta/2 + cv_r}{2c^2}, \; -\frac{(\gamma-1)v_r + c}{2c^2}, \; -\frac{(\gamma-1)v_\theta}{2c^2}, \; -\frac{(\gamma-1)v_z}{2c^2}, \; \frac{(\gamma-1)}{2c^2} \right),$$

$$L_2 = (-v_\theta, \; 0, \; 1, \; 0, \; 0),$$

$$L_3 = \left( 1 - \frac{(\gamma-1)\Theta}{2c^2}, \; \frac{(\gamma-1)v_r}{c^2}, \; \frac{(\gamma-1)v_\theta}{c^2}, \; \frac{(\gamma-1)v_z}{c^2}, \; -\frac{\gamma-1}{c^2} \right), \quad (A.7)$$

$$L_4 = (-v_z, \; 0, \; 0, \; 1, \; 0),$$

$$L_5 = \left( \frac{(\gamma-1)\Theta/2 - cv_r}{2c^2}, \; -\frac{(\gamma-1)v_r - c}{2c^2}, \; -\frac{(\gamma-1)v_\theta}{2c^2}, \; -\frac{(\gamma-1)v_z}{2c^2}, \; \frac{(\gamma-1)}{2c^2} \right).$$

In the TVD scheme which is based on the Eulerian grid, while physical quantities are defined in the grid center, $(i,j)$, fluxes are computed on the grid boundary, $(i+\frac{1}{2},j)$ and $(i,j+\frac{1}{2})$. We use the Roe's approximation (Roe 1981) to get the averaged values of the physical quantities at the grid boundary $(i+\frac{1}{2},j)$:

$$v_{x,i+\frac{1}{2},j} = \frac{\sqrt{\rho_{i,j}}v_{x,i,j} + \sqrt{\rho_{i+1,j}}v_{x,i+1,j}}{\sqrt{\rho_{i,j}} + \sqrt{\rho_{i+1,j}}},$$

$$v_{y,i+\frac{1}{2},j} = \frac{\sqrt{\rho_{i,j}}v_{y,i,j} + \sqrt{\rho_{i+1,j}}v_{y,i+1,j}}{\sqrt{\rho_{i,j}} + \sqrt{\rho_{i+1,j}}},$$

$$v_{z,i+\frac{1}{2},j} = \frac{\sqrt{\rho_{i,j}}v_{z,i,j} + \sqrt{\rho_{i+1,j}}v_{z,i+1,j}}{\sqrt{\rho_{i,j}} + \sqrt{\rho_{i+1,j}}}, \quad (A.8)$$

$$H_{i+\frac{1}{2},j} = \frac{\sqrt{\rho_{i,j}}H_{i,j} + \sqrt{\rho_{i+1,j}}H_{i+1,j}}{\sqrt{\rho_{i,j}} + \sqrt{\rho_{i+1,j}}},$$

$$c_{i+\frac{1}{2},j} = \left[ (\gamma-1)\left\{ H_{i+\frac{1}{2},j} - \frac{1}{2}(v^2_{x,i+\frac{1}{2},j} + v^2_{y,i+\frac{1}{2},j} + v^2_{z,i+\frac{1}{2},j}) \right\} \right]^{1/2}.$$

Similarly, we get the averaged values at $(i, j+\frac{1}{2})$.

In updating the state vector $\boldsymbol{q}^n$ to $\boldsymbol{q}^{n+1}$ (here the superscript $n$ is the time step), we calculate the fluxes along the $r$ and $z$ directions separately by the Strang-type dimensional splitting (Strang 1968) in the *hydrodynamic* step and take care of the source later in the *source* step. However, before the hydrodynamic step, we define an updated state vector which includes the effects of the source terms and the *curvature* terms at half time step

$$\boldsymbol{q}^{n*} = \boldsymbol{q}^n + \frac{\Delta t^n}{2}\left[ \boldsymbol{S}(\boldsymbol{q}^n) - \frac{\boldsymbol{F}_1(\boldsymbol{q}^n)}{r} \right], \quad (A.9)$$

and use the updated state vector in calculating fluxes. This *preprocessing* step will insure the formal accuracy of the code to be of second order.

In the hydrodynamic step, we calculate the $r$-flux as follows:

$$\boldsymbol{q}^{nr}_i = \boldsymbol{q}^n_i - \frac{\Delta t^n}{\Delta r}\left[ \frac{1}{r_i}\left( r_{i+\frac{1}{2}}\bar{\boldsymbol{f}}^*_{1,i+\frac{1}{2}} - r_{i-\frac{1}{2}}\bar{\boldsymbol{f}}^*_{1,i-\frac{1}{2}} \right) + \left( \bar{\boldsymbol{f}}^*_{2,i+\frac{1}{2}} - \bar{\boldsymbol{f}}^*_{2,i-\frac{1}{2}} \right) \right], \quad (A.10)$$

$$\bar{\boldsymbol{f}}^*_{1,i+\frac{1}{2}} = \frac{1}{2}\left[ \boldsymbol{F}_1(\boldsymbol{q}^{n*}_i) + \boldsymbol{F}_1(\boldsymbol{q}^{n*}_{i+1}) \right] - \frac{\Delta r}{2\Delta t^n}\sum_{k=1}^{5}\beta^*_{k,i+\frac{1}{2}}\boldsymbol{R}^{n*}_{k,i+\frac{1}{2}}, \quad (A.11)$$

$$\bar{\boldsymbol{f}}^*_{2,i+\frac{1}{2}} = \frac{1}{2}\left[ \boldsymbol{F}_2(\boldsymbol{q}^{n*}_i) + \boldsymbol{F}_2(\boldsymbol{q}^{n*}_{i+1}) \right], \quad (A.12)$$

$$\beta^*_{k,i+\frac{1}{2}} = Q_k \left( \frac{\Delta t^n}{\Delta r} a^{n*}_{k,i+\frac{1}{2}} + \gamma^*_{k,i+\frac{1}{2}} \right) \alpha^*_{k,i+\frac{1}{2}} - (g^*_{k,i} + g^*_{k,i+1}), \tag{A.13}$$

$$\alpha^*_{k,i+\frac{1}{2}} = \boldsymbol{L}^{n*}_{k,i+\frac{1}{2}} \cdot (\boldsymbol{q}^{n*}_{i+1} - \boldsymbol{q}^{n*}_{i-1}), \tag{A.14}$$

$$\gamma^*_{k,i+\frac{1}{2}} = \begin{cases} (g^*_{k,i+1} - g^*_{k,i})/\alpha^*_{k,i+\frac{1}{2}}, & \text{for } \alpha^*_{k,i+\frac{1}{2}} \neq 0 \\ 0, & \text{for } \alpha^*_{k,i+\frac{1}{2}} = 0 \end{cases}, \tag{A.15}$$

$$g^*_{k,i} = \text{sign}(\hat{g}^*_{k,i+\frac{1}{2}}) \max\left[0, \min\left\{|\hat{g}^*_{k,i+\frac{1}{2}}|, \hat{g}^*_{k,i-\frac{1}{2}} \text{sign}(\hat{g}^*_{k,i+\frac{1}{2}})\right\}\right], \tag{A.16}$$

$$\hat{g}^*_{k,i+\frac{1}{2}} = \frac{1}{2} \left[ Q_k(\frac{\Delta t^n}{\Delta r} a^{n*}_{k,i+\frac{1}{2}}) - (\frac{\Delta t^n}{\Delta r} a^{n*}_{k,i+\frac{1}{2}})^2 \right] \alpha^*_{k,i+\frac{1}{2}}, \tag{A.17}$$

$$Q_k(x) = \begin{cases} \left[x^2/(4\varepsilon)\right] + \varepsilon, & \text{for } |x| < 2\varepsilon \\ |x|, & \text{for } |x| \geq 2\varepsilon \end{cases}, \quad \varepsilon = \begin{cases} 0.1, & \text{for } k = 1 \text{ and } 5 \\ 0, & \text{for } k = 2, 3, \text{ and } 4 \end{cases}, \tag{A.18}$$

where the quantities with the superscript $*$ are computed with $\boldsymbol{q}^{n*}$. The subscript $i$ is the cell label along the $r$-axis, and the subscript for the $z$-axis is omitted for simplicity. Similarly, we calculate the $z$-flux with the state vector updated with the $r$-flux

$$\boldsymbol{q}^{nrz}_j = \boldsymbol{q}^{nr}_j - \frac{\Delta t^n}{\Delta z} \left( \bar{\boldsymbol{g}}^r_{j+\frac{1}{2}} - \bar{\boldsymbol{g}}^r_{j-\frac{1}{2}} \right), \tag{A.19}$$

$$\bar{\boldsymbol{g}}^r_{j+\frac{1}{2}} = \frac{1}{2} \left[ G(\boldsymbol{q}^{nr}_j) + G(\boldsymbol{q}^{nr}_{j+1}) \right] - \frac{\Delta z}{2\Delta t^n} \sum_{k=1}^{5} \beta^r_{k,j+\frac{1}{2}} \boldsymbol{R}^{nr}_{k,j+\frac{1}{2}}, \tag{A.20}$$

where $\beta$'s and $\boldsymbol{R}$'s are computed using the eigenvalues and eigenvectors of $\boldsymbol{B}$ with $\boldsymbol{q}^{nr}_j$. Then, the *hydrodynamically* updated state vector is $\boldsymbol{q}^n_{\text{hydro}} = \boldsymbol{q}^{nrz}$. To get a second order accuracy, the Strang-type dimensional splitting requires the order of the flux calculations to be permuted so that in the next time step the $z$-flux is calculated first and then $r$-flux, $\boldsymbol{q}^{n+1}_{\text{hydro}} = \boldsymbol{q}^{(n+1)zr}$.

Finally, in the source step, we include the source vector as follows:

$$\boldsymbol{q}^{n+1} = \boldsymbol{q}^n_{\text{hydro}} + \Delta t^n \boldsymbol{S}(\bar{\boldsymbol{q}}^{n+\frac{1}{2}}), \tag{A.21}$$

where

$$\bar{\boldsymbol{q}}^{n+\frac{1}{2}} = \frac{\boldsymbol{q}^n + \boldsymbol{q}^n_{\text{hydro}}}{2} \tag{A.22}$$

In order to get a true second order scheme and make the scheme stable even for the large source terms, the source terms should be included implicitly. But we use the explicit scheme instead, because the implicit scheme involves quite complication and the source terms are not too large to make any noticeable difference between the implicit and explicit schemes.

The time step $\Delta t^n$ is restricted by the usual Courant condition for the stability, and we calculate

$$\Delta t^n = C_{\text{courant}} \min \left[ \frac{\Delta r}{\max \left( |v^n_{r,i+\frac{1}{2},j}| + c^n_{i+\frac{1}{2},j} \right)}, \frac{\Delta z}{\max \left( |v^n_{z,i,j+\frac{1}{2}}| + c^n_{i,j+\frac{1}{2}} \right)} \right] \quad (A.23)$$

with $C_{\text{courant}} = 0.8$.

The Harten's TVD scheme was originally designed to capture shocks accurately. Tests showing the performance of shock capturing are found in the original paper by Harten (1983) for one-dimensional shocks and Ryu *et al.*(1993) for multi-dimensional shocks. Shocks in the TVD code are typically resolved within $2 - 3$ cells.

Our code runs at the speed of $\sim 8 \times 10^3$ zones per second with one CPU on a Convex 220 or at $\sim 5 \times 10^4$ zones per second with one CPU on a Cray 2.

# FIGURE CAPTIONS

Fig. 1.— Temporal evolution of the positive (kinetic plus internal) energy and the bremsstrahlung emission in (a) Run 1, (b) Run 2, (c) Run 3, and (d) Run 4. Both quantities are averaged over the computational box.

Fig. 2.— Velocity field at the end of the simulation, $t = 180$, in Run 4. The full $r$-$z$ plane with $-4 \leq r \leq 4$ and $-4 \leq z \leq 4$ is shown, although the simulation was performed on the first quadrant with $0 \leq r \leq 4$ and $0 \leq z \leq 4$. Solid lines in the first quadrant are the approximate trajectories of the fluid elements that enter through the boundary. Dotted lines are the approximate trajectories of fluid elements which are trapped in the vortices. The dashed lines indicate the two shocks which are identified with $\nabla \cdot \boldsymbol{v} < 0$.

Fig. 3.— Gray scale images of (a) density, (b) pressure, (c) entropy, and (d) Mach number at the end of the simulation, $t = 180$, in Run 4. Black indicates the regions with high (or positive) values and white those with low (or negative) values. The full $r$-$z$ plane with $-4 \leq r \leq 4$ and $-4 \leq z \leq 4$ is shown.

Fig. 4.— Velocity field at the end of the simulation, $t = 180$, (a) in Run 1, (b) in Run 2, (c) in Run 3, and (d) in Run 4. The full $r$-$z$ plane with $-4 \leq r \leq 4$ and $-4 \leq z \leq 4$ is shown. Solid lines show the funnel wall according to equation (2.13).

Fig. 5.— The bremsstrahlung emission ($\dot{E}_{\rm brem}$), the angle of the conical outflowing wind ($\theta_{\rm out}$), the shock stand-off distance in the equatorial plane ($r_s$), and the mass within the shock in the trapped annulus ($m_A$) in the calculations Run 1 - 4 with different values of the dimensionless parameter $\epsilon$.

Fig. 6.— Temporal evolution of the positive (kinetic plus internal) energy and the bremsstrahlung emission in the simulations for the accretion flow with vanishing inflow thickness. Both quantities are averaged over the computational box.

Fig. 7.— Velocity field at $t = 66, 67, 68,$ and $69$ (during quiet phase) in the simulations for the accretion flow with vanishing inflow thickness. The $x$-$z$ plane with $-0.2 \leq x \leq 0.6$ and $0. \leq z \leq 0.8$ is shown, although the simulation was performed with $-0.4 \leq x \leq 0.6$ and $0 \leq z \leq 1$.

Fig. 8.— Velocity field at $t = 86, 87, 88,$ and $89$ (during active phase) in the simulations for the accretion flow with vanishing inflow thickness. The $x$-$z$ plane with $-0.2 \leq x \leq 0.6$ and $0. \leq z \leq 0.8$ is shown, although the simulation was performed with $-0.4 \leq x \leq 0.6$ and $0 \leq z \leq 1$.

Fig. 9.— Radial Mach number ($M_r$) and pressure ($p$) distributions of an one dimensional steady isentropic flow in cylindrical geometry. The flow enters through a boundary located at $r^* = 4$ with $\rho^* = 1$, $v_\theta^* = 1/4$, $v_z^* = 0$, $M_r^* = 10$, and $\gamma = 5/3$, same as those of Run 1 - Run 4 described in §4.

Fig. 10.— Comparison of the pressure distributions of one-dimensional steady isentropic flows in cylindrical geometry with shocks (a) at $r = 2.4$ and $r = 2.5$ and (b) at $r = 0.9$ and $r = 1.0$. The flows enter through a boundary located at $r^* = 4$ with $\rho^* = 1$, $v_\theta^* = 1/4$, $v_z^* = 0$, $M_r^* = 10$, and $\gamma = 5/3$, same as those of Run 1 - Run 4 described in §4.

Fig. 11.— Power spectrum of the bremsstrahlung emission in Fig. 6 ($\dot{E}_{\rm brem}/h_{\rm in}^2$) in the time interval $25 \leq t \leq 100$ as a function of frequency (period inverse).